\documentclass{article}
\usepackage{spconf,amsmath,graphicx}
\usepackage{subfig}
\usepackage{tikz}
\usepackage{url}
\usepackage{pifont}
\newcommand{\cmark}{\ding{51}}
\newcommand{\xmark}{\ding{55}}

\title{IMPROVING REVERBERANT SPEECH SEPARATION WITH MULTI-STAGE TRAINING AND CURRICULUM LEARNING}
\name{Rohith Aralikatti, Anton Ratnarajah, Zhenyu Tang, Dinesh Manocha}
\address{University of Maryland, College Park\\
\{rohithca, jeran, zhy, dmanocha\}@umd.edu}

\begin{document}
%
\maketitle
\begin{abstract}

We present a novel approach that improves the performance of reverberant speech separation. Our approach is based on an accurate geometric acoustic simulator (GAS) which generates realistic room impulse responses (RIRs) by modeling both specular and diffuse reflections. We also propose three training methods - pre-training, multi-stage training and curriculum learning that significantly improve separation quality in the presence of reverberation. We also demonstrate that mixing the synthetic RIRs with a small number  of real RIRs during training enhances separation performance. We evaluate our approach on reverberant mixtures generated from real, recorded data (in several different room configurations) from the VOiCES dataset. Our novel approach (curriculum learning+pre-training+multi-stage training) results in a significant relative improvement over prior techniques based on image source method (ISM).

\end{abstract}
\section{Introduction and Related Work}
\label{sec:intro}

Speech separation (also referred to as the cocktail party problem) has been well-studied for several decades. Humans are able to easily separate multiple streams of audio/speech because they use both spatial audio information as well as higher contextual cues (such as the content of what is being spoken). However, state-of-the-art algorithms or models still struggle to match human performance, especially in the presence of noise and reverberation.
The earliest approaches to speech separation were based on signal processing techniques such as computational auditory scene analysis (CASA) \cite{cooke2001auditory}, non-negative matrix factorization (NMF) \cite{schmidt2007speech} and independent component analysis (ICA) \cite{hyvarinen2000independent}.

With recent advances in deep learning, supervised approaches to speech separation have become popular. In such approaches, CNN/LSTM models are trained on large datasets to learn the mapping between the speech mixture and individual components. Some well known deep learning methods for speech separation include deep clustering \cite{hershey2016deep} and permutation invariant training (PIT) \cite{yu2017permutation}.

In real-world scenarios, speech signals are reflected from the walls and objects present in the room to create reverberation effects. Most well-studied algorithms and models for source separation do not adequately account for  such reverberation effects. As a result, this can impact the real-world performance of speech separation systems.
Recently, there have been several works in reverberant speech separation. In \cite{maciejewski2020whamr}, the authors release the WHAMR! dataset which is a noisy reverberant augmentation of the wsj-mix \cite{hershey2016deep} dataset. The authors present a multi-stage dereverberation, denoising and separation network, along with baseline speech separation results on the WHAMR! dataset.
The  Wavesplit architecture~\cite{zeghidour2020wavesplit} demonstrates state of the art results on the WHAMR! dataset. Auxiliary autoencoding training (A2T)~\cite{luo2021distortion} has also been shown to improve reverberant speech separation. DBNet~\cite{aroudi2021dbnet} is a single network that can be used for direction-of-arrival estimation, beam-forming and reverberant speech enhancement. In \cite{zhang2021time}, the authors propose a multi-channel time domain speech extraction network that uses speaker embeddings to extract speech from a specific speaker, given a reverberant and noisy input mixture. In most existing work on reverberant speech separation, the models are trained and tested on synthetically generated datasets. 
As a result, their accuracy is governed by these datasets.
Other techniques focus on improving reverberant speech separation performance by modifying the underlying model architecture. 

In addition to generation of synthetic data, there is some work on using different learning methods to improve the training performance. In our approach, we make use of
curriculum learning~\cite{bengio2009curriculum}. It corresponds to a training procedure, where the model is first trained on simpler data and the model is slowly exposed to more complex data as the training progresses. Curriculum learning has been shown to be helpful in many tasks such as automatic speech recognition (ASR) \cite{braun2017curriculum}, far-field ASR \cite{ranjan2021curriculum}, speech translation \cite{wang2020curriculum}, audiovisual speech recognition \cite{afouras2018deep} and audiovisual speech enhancement \cite{afouras2018conversation}. In \cite{braun2017curriculum}, curriculum learning is used to make the speech recognition model more robust to ambient noise. 
This is performed by first training the model with speech samples at high signal-to-noise ratio (SNR) of 50 dB and gradually reducing the SNR to 0 dB as training progresses.
In \cite{ranjan2021curriculum}, the authors  attempt to improve the performance of far-field ASR by applying a curriculum which increases the distance between the source and the listener, while the training progresses. The authors show that slowly progressing from training the model on near-field utterances to far-field utterances improves far-field ASR performance. 
For audio-visual ASR and speech enhancement, curriculum learning has typically been applied as a pre-training step to ensure that the model does not ignore the visual modality in favour of the audio modality.
(as the audio modality is usually a richer source of information for such tasks). 
In such cases, the visual model is first pre-trained on tasks of slowly increasing difficulty before the combined audio-visual model is trained together for the specified task. 
To the best of our knowledge, such curriculum learning techniques have not been applied to the task of reverberant speech separation.

Pre-training is another well-known machine learning technique that has been shown to improve model robustness and uncertainty estimates \cite{hendrycks2019using}. In speech processing, pre-training techniques have been used for a variety of applications. Pre-training has been utilized to train low resource ASR systems using data from a high resource language \cite{bansal2018pre}. Pre-training has also been used to improve the rate of convergence and the word-error-rate (WER) in the case of end-to-end training of automatic speech recognition (ASR) systems \cite{zeyer2018improved}. Unsupervised pre-training has also been shown to improve the accuracy of ASR systems \cite{schneider2019wav2vec}. Pre-training has also been shown to reduce label permutation instability while training speech separation models \cite{huang2020self}. In this work, we show that a simple pre-training method causes a significant improvement in the performance of reverberant speech separation.

\textbf{Main Contributions:} 
We present new techniques to train a model using a mixture of synthetic and real data. We use more accurate methods to improve separation quality by focusing on the quality of synthetic RIR generation methods used during training. We also propose novel training procedures which seek to make better use of synthetic data to improve the real-world performance.
\begin{itemize}
  \item We present a novel approach that augments training data by using synthetic RIRs generated by an geometric acoustic simulator that accounts for specular and diffuse reflections. In particular, higher order diffuse reflections can accurately model the reverberation effects. Our formulation improves reverberant speech separation performance as compared to the synthetic  RIRs generated by an image source method (ISM)~ \cite{allen1979image,diaz2021gpurir}.  We show that an accurate geometric acoustic simulator (GAS) shows an average relative improvement of $21\%$ in SI-SDRi as compared to the traditional image source method.
  \item We present improved training techniques based on curriculum learning~\cite{bengio2009curriculum}, multi-stage training and pre-training to improve the overall performance. We show that pre-training the model on non-reverberant speech separation data can significantly improve real-world separation performance. We show that training techniques such as pre-training and curriculum learning result in a relative improvement of $43\%$ in SI-SDRi, when compared to the baseline. 
\end{itemize}

In section \ref{sec:dataset}, we present the details of the datasets used in our approach. Section~\ref{sec:rirgen} describes our approach to generate synthetic RIRs. Section \ref{sec:experiments} presents details about the model architecture and the model training parameters used in our experiments.

\section{DATASETS}
\label{sec:dataset}

The LibriMix dataset \cite{cosentino2020librimix} is an open-source speech separation dataset based on the LibriSpeech Corpus \cite{panayotov2015librispeech}. All experiments in this paper are run on the 100-hour split of Libri2Mix data (the two speaker mixture subset of the LibriMix dataset) at a sampling rate of 8000 hertz. The "min" criterion is used for mixing speech signals; that is, the duration of the mixed speech signal is set to be the minimum of the duration of the two speech signals being mixed. There are 13,000 data samples in the training split and 3,000 data samples each in the testing and validation splits. All experiments in this paper are done on the two speaker subset of LibriMix data (Libri2Mix).

The LibriMix dataset is a non-reverberant speech separation dataset. Reverberation is introduced on the fly during training by convolving the separated speech signals with a randomly chosen RIR (can be either real or synthetic) before mixing. We do not employ dynamic mixing \cite{zeghidour2020wavesplit} while generating speech mixtures. 

Real RIRs are obtained from the BUTReverb RIR dataset \cite{szoke2019building}. We split the real RIRs into train (1171 RIRs), dev (251 RIRs) and test (251 RIRs) subsets. The same splits are used to generate all experimental results. Synthetic RIRs are generated as specified in section \ref{sec:rirgen}.

All the models were trained on the reverberant LibriMix dataset and then tested on recordings from the VOiCES dataset \cite{richey2018voices}. This was done to ensure that we have one common test set consisting of real reverberant mixtures on which the performance of different approaches can be compared. The VOiCES dataset consists of four different room configurations in which data has been recorded. For each room configuration, we generate 400 test mixtures by randomly mixing two recordings. The room configurations used to record VOiCES data is mentioned in Table \ref{tab:room-dimensions-VOiCES}.

\section{RIR generation and Training}
\label{sec:rirgen}

\begin{table}
\begin{center}
\caption{Room dimensions (in meters) used to generate synthetic RIRs using GAS and gpuRIR. These are the same room dimensions used to capture real RIR data in the BUTReverb dataset \cite{szoke2019building}.}


 \begin{tabular}[htb!]{||c c c||} 
 \hline
 \textbf{Length} & \textbf{Width} & \textbf{Height} \\ [0.5ex] 
 \hline\hline
 10.7 & 6.9 & 2.6 \\ 
 \hline
 4.6 & 6.9 & 3.1  \\
 \hline
 7.5 & 4.6 & 3.1  \\
 \hline
 6.2 & 2.6 & 14.2  \\
 \hline
  28.1 & 11.1 & 3.3  \\
 \hline
 11.5 & 20.1 & 4.8  \\
 \hline
 17.2 & 22.8 & 6.9  \\
 \hline
  7.0 & 4.1 & 3.6  \\
 \hline
  4.4 & 2.8 & 2.7  \\ 
 \hline
\end{tabular}
\label{tab:room-dimensions}
\vspace*{-0.14in}

\end{center}
\end{table}

Sections \ref{subsec:gpuRIR}, \ref{subsec:GAS}, \ref{subsec:IRGAN} and \ref{subsec:TS-RIRGAN} describe the methods used to generate synthetic RIRs used in our experiments. Sections \ref{ssec:curr}, \ref{ssec:PT} and \ref{ssec:multi} explain the novel training methods employed to improve the performance of our speech separation model.

For both RIR generation methods described below, room dimensions as specified in Table \ref{tab:room-dimensions} were used to generate RIRs. These room dimensions are the same room dimensions used to capture real RIRs in the BUTReverb dataset. This was done to ensure consistency between synthetic and real RIRs used during training.

\begin{figure*}[t] 
\centering
\subfloat[Real RIR from VOiCES dataset]{\includegraphics[width=0.6\columnwidth]{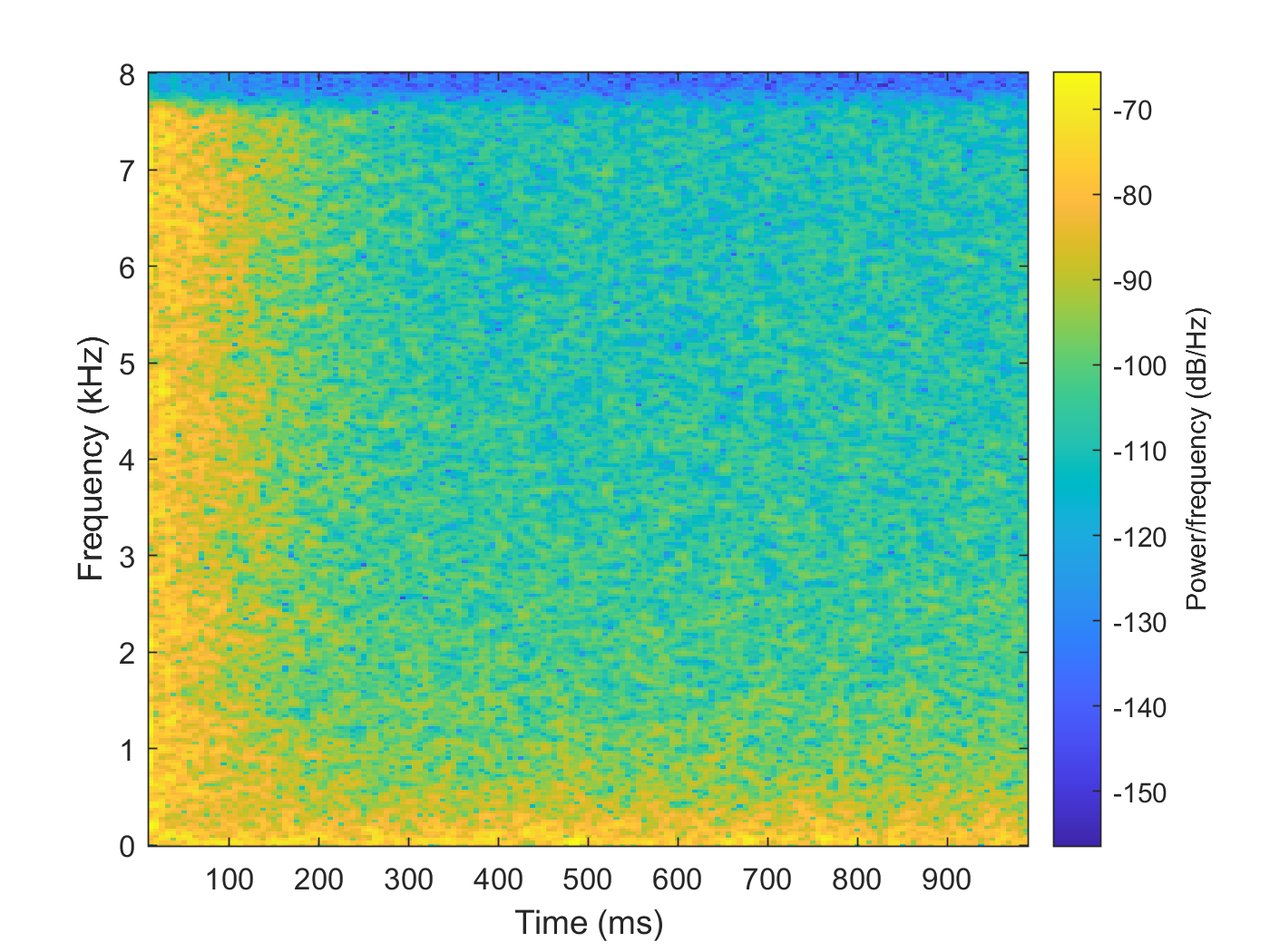} \label{GAS}}
\quad
\subfloat[GAS (specular + diffuse)]{\includegraphics[width=0.6\columnwidth]{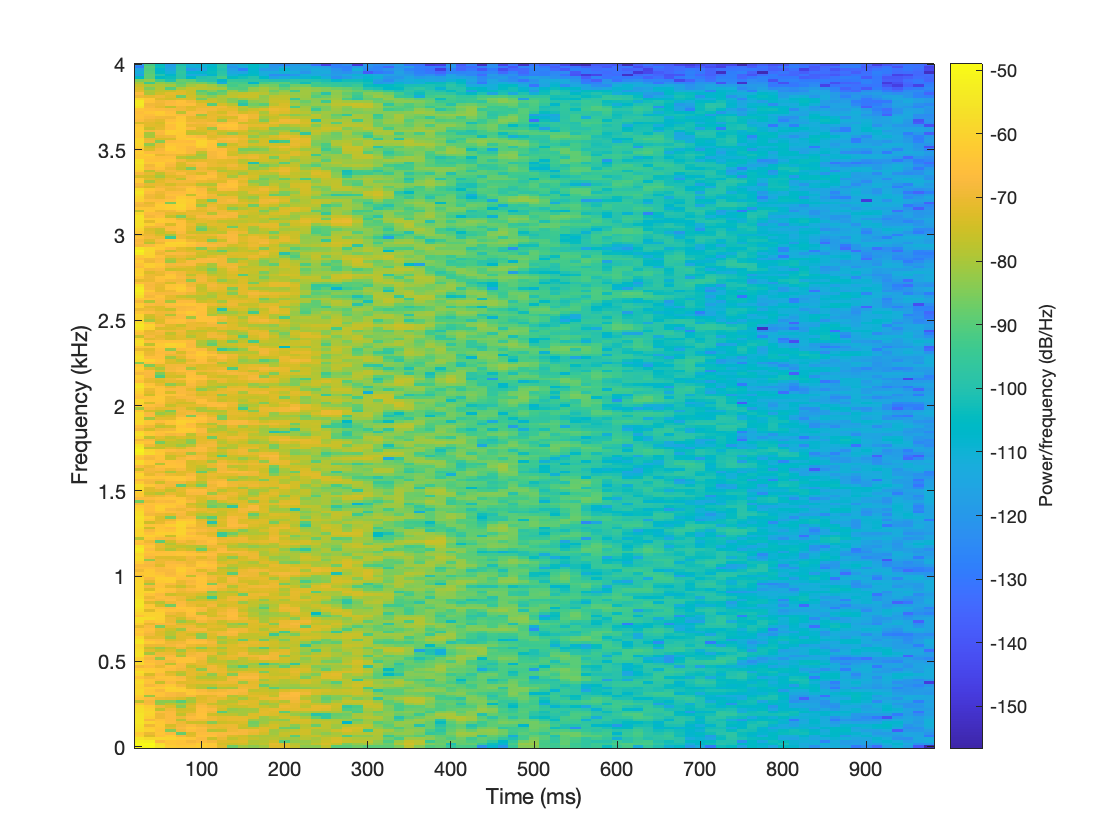}\label{GAS+EQ}}
\quad
\subfloat[gpuRIR(image source)]{\includegraphics[width=0.6\columnwidth]{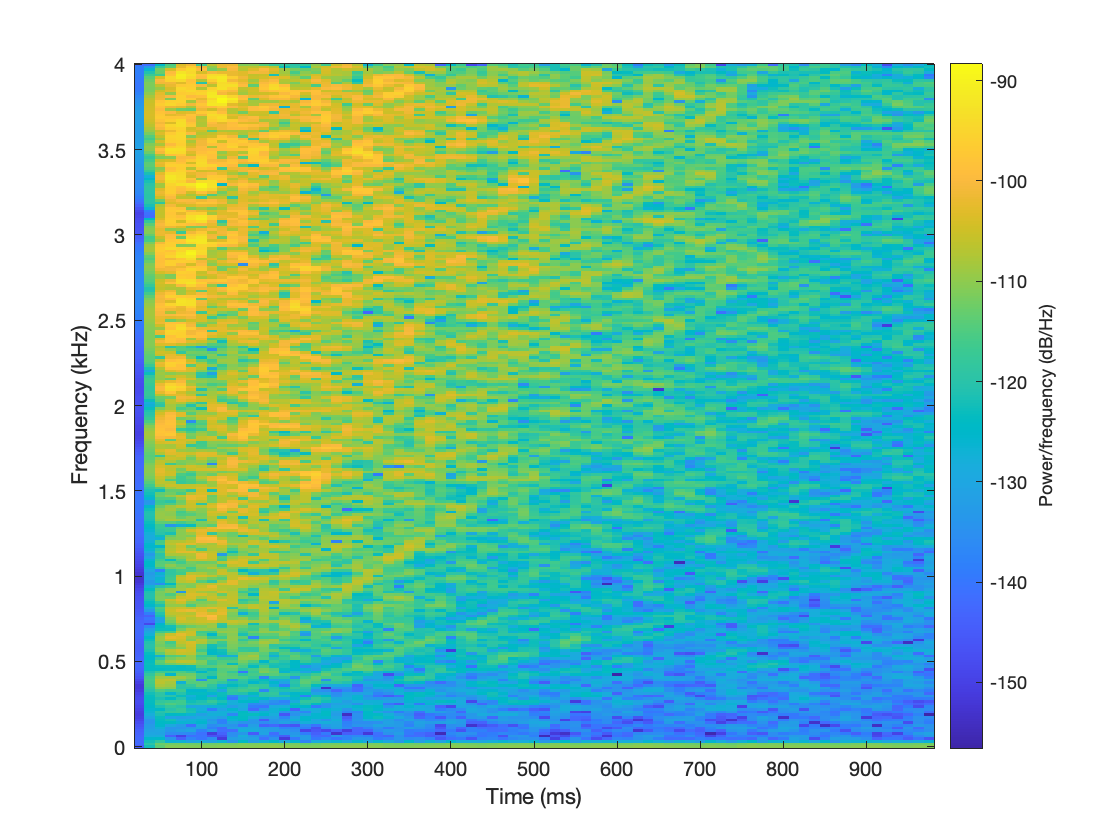}\label{T(GAS+EQ)}}
\quad
\subfloat[IR-GAN]{\includegraphics[width=0.6\columnwidth]{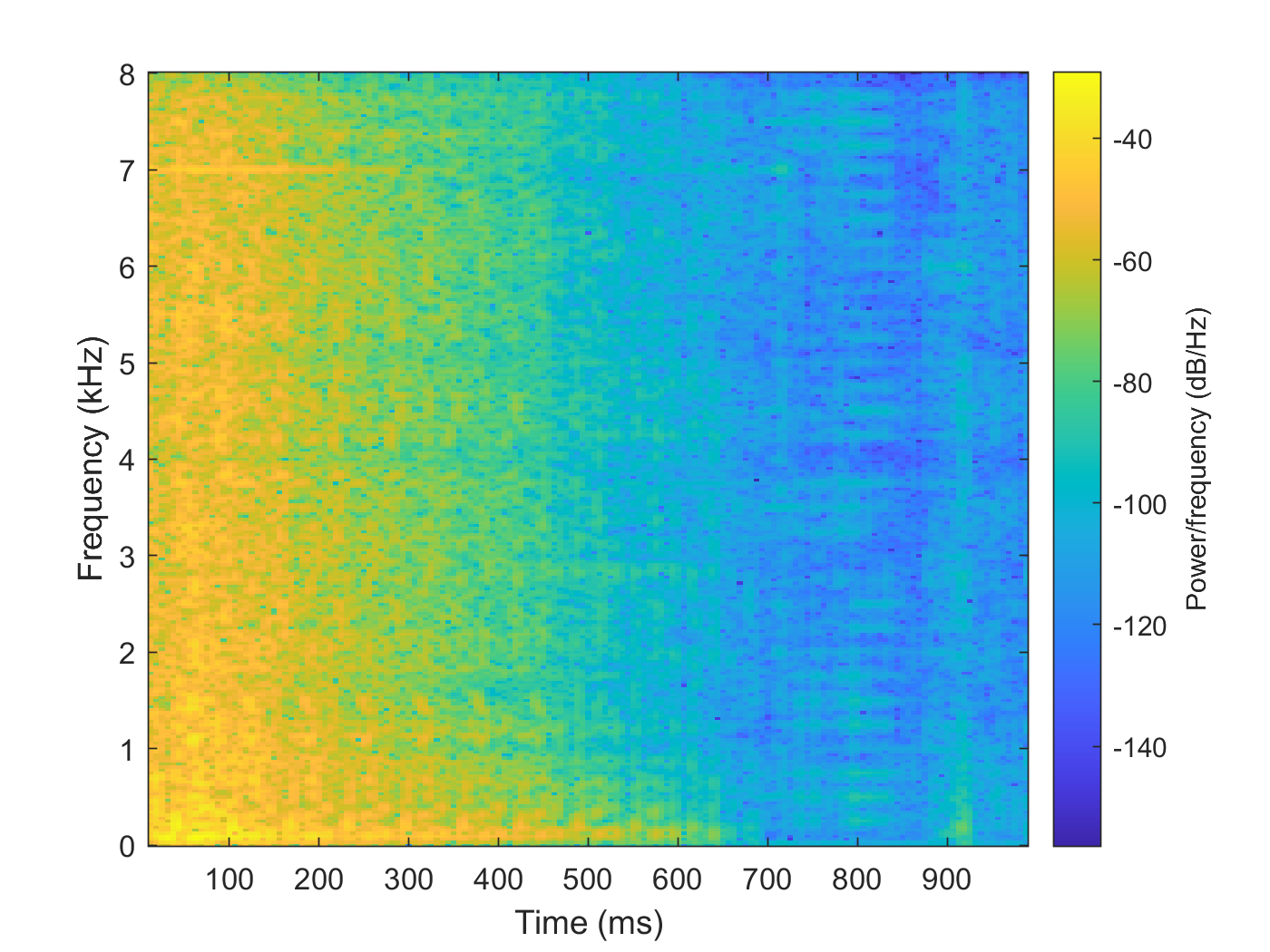}\label{T(GAS+EQ)}}
\quad
\subfloat[TS-RIRGAN (GAS+CycleGAN translation)]{\includegraphics[width=0.6\columnwidth]{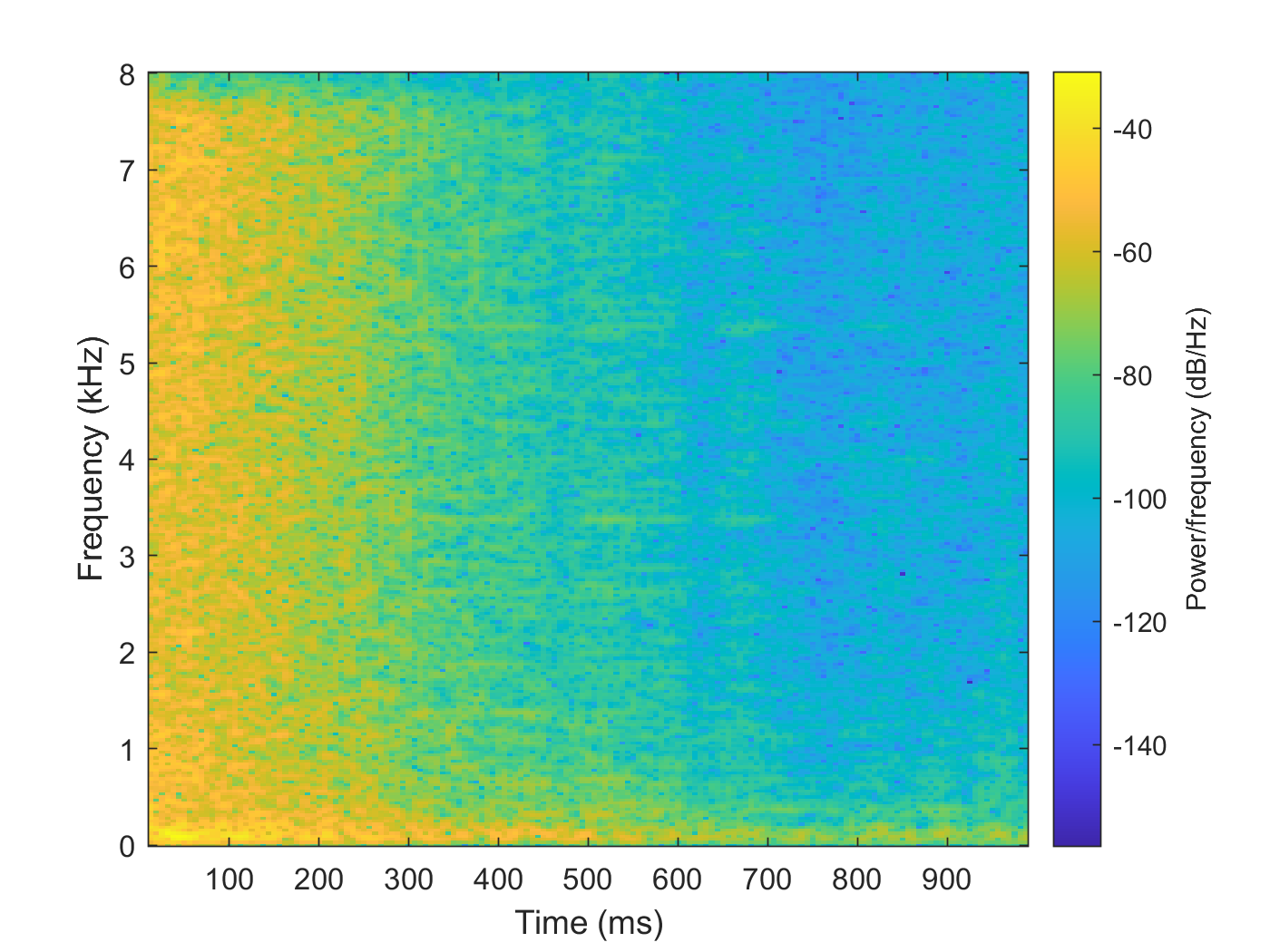}\label{T(GAS+EQ)}}
\caption{Comparison of the spectrograms of synthetic RIRs with a real RIR from the VOiCES dataset. Our approach based on GAS and GAN based generation methods are more accurate as compared to image source method used in gpuRIR.} 
\label{fig:1}
\end{figure*}

\subsection{gpuRIR}
\label{subsec:gpuRIR}

gpuRIR\cite{diaz2021gpurir} implements the image source method \cite{allen1979image} to generate RIRs. The Image Source Method (ISM) is one of the widely used algorithm techniques to calculate acoustic Room Impulse Responses (RIRs). However, it only models the specular reflections. The image method is the current most widely used method in the speech community for generating RIRs in various learning-based tasks~\cite{ko2017study}.  It assumes that all reflection paths can be constructed by mirroring sound sources with respect to the reflecting plane. As a result, a source will be mirrored multiple times depending on the desired order of reflections. In many cases, the image method fails to model the late reverberation part of an RIR.
Moreover, its computational complexity grows considerably with the order of reflections.
gpuRIR improves the computation speed of the ISM by using the parallel capabilities of Graphic Processing Units for accelerating the computation of images inside each RIR nd computing multiple RIRs.
The $beta\_SabineEstimation$ function provided by gpuRIR can be used to estimate room absorption coefficients to generate RIRs of desired $T_{60}$.

\subsection{Geometric Acoustic Simulator (GAS)}
\label{subsec:GAS}

A key aspect of our approach is to generate synthetic RIRs that model diffuse effects.
Diffuse reflections occur when sound energy is scattered into non-specular directions. Most real world scenarios consist of diffuse reflections and they are widely studied in room acoustics~\cite{hodgson1991evidence,dalenback1994macroscopic}. There is considerable work on accurately simulating diffuse effects using  
stochastic path tracing. In these methods, the sound paths are randomly traced in all directions and each path follows either specular or diffuse reflections. The resulting path tracers use a  scattering coefficient, which is typically between $0$ and $1$. This scattering coefficient denotes the proportion of sound energy that is diffusely reflected at a surface ($0$ means perfectly specular and $1$ means perfectly diffuse).
These geometric techniques model the acoustic effects based on ray theory and typically work well for high-frequency sounds to model specular as well as diffuse reflections~\cite{krokstad1968,lauterbach2007,savioja2015overview}. Our approach is based on a geometric acoustic simulator (GAS)~\cite{schissler2014high,schissler2018interactive}, which models occlusions and specular and diffuse reflections. On the other hand, image source method only models specular reflections. In practice, GAS can generate synthetic RIRs that are closer to real-world RIRs, as shown in Fig.\ref{fig:1}.
When combined with real RIRs, synthetic IRs generated using GAS have been shown to improve the performance of speech recognition as well as keyword spotting tasks~\cite{tang2020improving}.

\subsection{IR-GAN}
\label{subsec:IRGAN}


IR-GAN \cite{ratnarajah2020ir} is a GAN based RIR generator that can generate realistic RIRs. IR-GAN has been shown to improve the performance of far-field speech recognition. IR-GAN learns acoustic features (e.g., reverberation time ($T_{60}$), direct to reverberation ratio (DRR) etc.) from real RIRs and can generate RIRs with different combination of acoustic features by considering different linear combinations of the noise vector used to generate RIRs.  The performance of IR-GAN depends on the real-world RIRs used to train the network. We train IR-GAN on real RIRs from the BUTReverb dataset in order to generate synthetic RIRs for our speech separation experiments.

\subsection{TS-RIRGAN}
\label{subsec:TS-RIRGAN}

TS-RIRGAN \cite{ratnarajah2021ts} is a CycleGAN model that learns an unpaired data mapping from synthetic RIRs to real RIRs. TS-RIRGAN improves the quality of synthetic RIRs by compensating low frequency wave effects. We used pre-trained TS-RIRGAN~\footnote{\url{https://github.com/anton-jeran/TS-RIR}} to improve the quality of synthetic RIRs generated using GAS. TS-RIRGAN model is trained using real RIRs from BUT ReverbDB dataset and synthetic RIRs from GAS. We translate the synthetic RIRs generated by GAS in section \ref{subsec:GAS} to generate improved synthetic RIRs. No further post-processing of these RIRs is done prior to training the speech separation model.

\subsection{Pre-training}
\label{ssec:PT}

We initialize the model weights used for the reverberant training runs with the weights of the converged model trained on non-reverberant speech separation data (the unmodified Libri2Mix dataset). The pre-trained model achieved an SI-SDRi of 13.9 and SDRi of 14.31 on the unmodified Libri2Mix dataset. The intuition behind such a pre-training step was that the model has already learned to separate speech signals from mixtures and must now learn to account only for the reverberation caused by repeated reflections from the walls and nearby objects present in the room. This is a slightly simpler problem to learn as compared to separating speech signals and handling reverberant effects in a single step. When pre-training and curriculum learning (described in the next section) are applied together, the model training becomes a gradual process where the speech separation model learns to handle increasing amounts of reverberation.

\subsection{Curriculum Learning}
\label{ssec:curr}

\begin{figure}[!htb]
    \centering
    \includegraphics[width=\columnwidth]{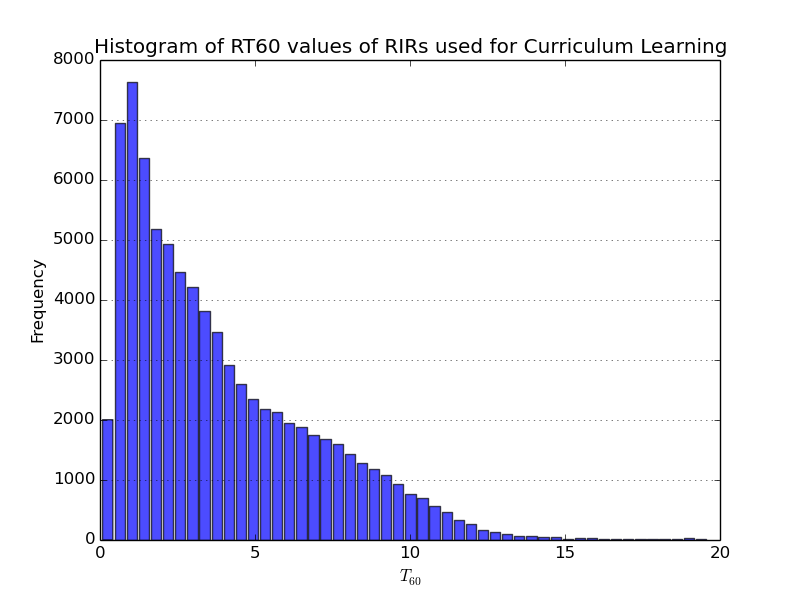}
    \caption{Histogram of $T_{60}$ values of RIRs used for curriculum learning.}
    \label{fig:hist_curr}
\end{figure}

The synthetic and real RIRs are arranged in ascending order of  $T_{60}$ values. At the start of the training, only those RIRs with low  $T_{60}$ values are chosen to generate reverberant data on which the model is trained. As training progresses, the  $T_{60}$ threshold is slowly increased over time and the model learns to separate mixtures with increasing reverberation in a gradual manner.
We generate 40k synthetic RIRs of differing $T_{60}$ values using GAS. This is done by varying the sound absorption coefficient from 0 to 1 in steps of 0.01. For each value of absorption coefficient, we generate 400 RIRs. The room dimensions used to generate these RIRs are specified in Table \ref{tab:room-dimensions}. The histogram of $T_{60}$ values of the RIRs generated for curriculum learning is shown in Figure \ref{fig:hist_curr}.
The RIRs are sorted in ascending order of $T_{60}$ values. At the start of the training, the model is initialized with weights from the pre-trained model. This pre-trained model performs well on the task of reverb-free speech separation. When training begins, the 400 RIRs with the lowest $T_{60}$ values are chosen to generate the reverberant signals from which the reverberant mixtures are obtained for training. From the histogram in Figure \ref{fig:hist_curr}, we see that the first 400 RIRs have $T_{60}$ of close to zero. Convolving speech signals with such RIRs produces mixtures with very low reverberation which the pre-trained model can easily learn to separate. After every 2nd epoch, we include the next set of 400 RIRs with slightly higher $T_{60}$ values in the set of available RIRs for training. Hence, at each stage the model slowly adjusts to increasing reverberation and training proceeds smoothly. Since we have 40k RIRs in total and we train for 200 epochs, at the end of training all the synthetic RIRs generated (including those with high $T_{60})$ will have been used for training the model.

\subsection{Multi-stage training}
\label{ssec:multi}

We propose a training pipeline where RIRs generated from the methods described in Sections \ref{subsec:GAS} - \ref{subsec:TS-RIRGAN} are used to train the speech separation model in multiple stages. After pre-training on anechoic data, the model is first trained on reverberant data generated using RIRs from GAS. 
The model is then fine-tuned using RIRs generated from GAN based methods - IR-GAN and TS-RIRGAN. The intuition is that GAS can generate RIRs with a wide range of reverberation parameters ($T_{60}$ and DRR) which is used for the initial training. The GAN based methods have been shown to generate realistic RIRs and are used to fine-tune the model to improve performance on real reverberant data. We show that training the model in multiple stages leads to better performance when compared to a single training stage (where RIRs from different RIR generation methods are combined to generate the reverberant training data).

\section{Experiments}
\label{sec:experiments}

\begin{table*}[htb!]
\centering
 \caption{Ablation results (SI-SDRi) that quantify the contribution of real RIR data, pre-training, multi-stage training and curriculum learning to improving model performance. Our overall approach based on (curriculum learning+pre-training+GAS) gives the best results.}
 \begin{tabular}{|| c|c| c | c  |c | c | c | c ||} 
 \hline
 RIR & Pre-training & Real RIRs & Curriculum & \multicolumn{4}{c}{SI-SDRi} \\ [0.5ex] 
 
  Generation & & used? & learning & Room 1 &  Room 2  &  Room 3 & Room 4  \\
  \hline\hline
  Non-reverb baseline & NA & NA & NA & -4.25 & -4.34 & -4.25 & -4.18 \\
  gpuRIR (ISM) & \xmark & \xmark & \xmark & 2.27 & 2.13 & 1.48 & 2.19 \\
  gpuRIR (ISM) & \cmark & \cmark & \xmark & 4.35 & 3.71 & 1.96 & 3.61 \\
  GAS & \xmark & \xmark & \xmark &  3.44 &  2.09 &  1.09 & 2.73 \\ 
  GAS & \xmark & \cmark & \xmark  & 4.47  & 3.00  & 1.77  & 3.31  \\
GAS & \cmark & \xmark & \xmark  & 5.53  & 4.04  & 2.28  & 4.28  \\
 GAS & \cmark & \cmark & \xmark  & 5.85  & 4.62 &  2.51 & 4.27  \\
 GAS & \cmark & \cmark & \cmark  & 5.51 &  5.15  & 2.51 &  4.8  \\  

GAS + IR-GAN  & \cmark & \cmark & \xmark  & 5.06 &  4.78  & 2.25 &  4.34  \\
  
GAS $\rightarrow$ IR-GAN & \cmark & \cmark & \xmark  & 5.85 &  5.52  & 3.27 &  5.06  \\
   
    
 
GAS + IR-GAN + TS-RIRGAN & \cmark & \cmark & \xmark  & 5.4 &  4.55  & 2.43 &  4.5  \\ 
 
GAS $\rightarrow$ IR-GAN $\rightarrow$ TS-RIRGAN & \cmark & \cmark & \xmark  & 5.93 &  \textbf{5.55}  & \textbf{3.41} &  \textbf{5.23} \\

GAS $\rightarrow$ IR-GAN $\rightarrow$ TS-RIRGAN & \cmark & \cmark & \cmark  & \textbf{5.96} &  5.27  & 2.98 &  5.16 \\
 
 \hline
 \end{tabular}
 \label{tab:ablation}
\end{table*}

 

\subsection{Model Architecture and Training Details}
\label{ssec:model_arc}

All experiments are run on the DPRNN-TasNet \cite{luo2020dual} model. The model has 8 DPRNN blocks with 512 units in the RNN cell state. The ADAM optimizer \cite{kingma2014adam} is used with an initial learning rate of $10^{-3}$ and no weight decay. The learning rate is halved every time the validation loss plateaus. We use early stopping with a delay of 10 epochs to terminate the training. A batch size of 16 is used for all experiments. The model is trained for a maximum of 200 epochs. The asteroid framework \cite{pariente2020asteroid} is used for all experiments in this paper.

Synthetic RIRs are generated as specified in Section \ref{sec:rirgen}. While the model is being trained, reverberant data is generated on the fly by convolving a randomly chosen real or synthetic RIR with the source signal. The reverberant mixture is then obtained by mixing the reverberant speech signals. For each of the RIR generation methods specified in Section \ref{sec:rirgen}, we combine real and synthetic RIRs and train the same DPRNNTasNet model as specified above.
The model is trained to generate the separated reverberant speech signals given a reverberant input mixture.

\subsection{Results}
\label{ssec:results}

We report the improvement in 
scale-invariant signal-to-distortion ratio (SI-SDRi) as the metric to compare the different training approaches. We use the mir\_eval \cite{raffel2014mir_eval} library to generate this metric. The improvements are computed by obtaining the difference in metric values for the input mixture and the predicted separated signals.
We report results separately for each of the room configurations present in the VOiCES dataset. The four room configurations are specified in Table \ref{tab:room-dimensions-VOiCES}.

In Table \ref{tab:ablation}, we have reported the SI-SDRi 
values for the models trained on Libri2Mix and tested on two speaker mixtures created from the VOiCES dataset.

The non-reverb baseline - that is, the model trained on clean, non-reverberant data does extremely poorly on the reverberant speech separation task.

As expected, we see that RIR generation method that employs the image source method (gpuRIR) has lower SI-SDRi. This is reasonable because the image source method just models only specular reflections. Since GAS simulates both specular and diffuse reflections, it performs significantly better than gpuRIR when tested on real data. GAS has an average relative improvement of $21\%$ in SI-SDRi compared to gpuRIR.

The results also show that fine-tuning the GAS model using RIRs generated from IR-GAN and TS-RIRGAN improves the average SI-SDRi from $4.49$ dB to $5.04$ dB. This is an average relative improvement of $12.24\%$ over all room configurations.

The training names which include a "+" sign indicate that the RIRs generated from different training methods were used in a single training step. For example, "GAS + IR-GAN" indicates that half the training data was generated using RIRs from GAS and the other half was generated using RIRs from IR-GAN. The training names which include a "$\rightarrow$" specifies a multi-stage training pipeline. For example, "GAS $\rightarrow$ IR-GAN" implies that the model was first trained on reverberant data generated from GAS RIRs and was then fine-tuned on data generated from IR-GAN RIRs. The results clearly show that a multi-stage training approach works much better as compared to single stage training where different RIR generation methods are used at once.


We obtain the best SI-SDRi value of $x$ (averaged over all rooms) when we employ a multi-stage training pipeline combining GAS and GAN based RIR generation methods along pre-training.


\begin{table}
\begin{center}
\caption{Room dimensions (in meters) used to capture VOiCES test data. Room 4 consists of two adjacent cuboidal rooms and both dimensions are specified.}

 \begin{tabular}[h]{|| c | c c c||} 
 \hline
 \textbf{Room} & \textbf{Length} & \textbf{Width} & \textbf{Height} \\ [0.5ex] 
 \hline\hline
 Room 1 & 3.7 & 2.7 & 2.7 \\ 
 \hline
 Room 2 & 5.7 & 4.01 &  2.77  \\
 \hline
 Room 3 & 7.62 & 7.62  & Not specified  \\
 \hline
 Room 4 & 9.75+3.05 & 4.87+3.05 & Not specified  \\
 \hline
 
 \end{tabular}
\label{tab:room-dimensions-VOiCES}

\end{center}
\end{table}

\subsection{Ablation Studies}
\label{ssec:ablation_results}

In Table \ref{tab:ablation}, we demonstrate the effects of three factors on the performance of reverberant speech separation. The three factors studied are - inclusion of real RIRs during training, applying curriculum learning and pre-training the model.

All results shown in Table \ref{tab:ablation} have been trained on the same model (DPRNNTasNet). The dataset used for the ablation experiments is the reverberant LibriMix generated by convolving the original LibriMix dataset with synthetic RIRs generated from GAS/gpuRIR and real RIRs from BUTReverb data. 
We observe the following benefits from the results in Table \ref{tab:ablation}:

\begin{itemize}
    \item The inclusion of real RIRs in the training process is important. We see that the addition of even a small quantity of real RIR data in the training process significantly enhances the real-world performance of the reverberant speech separation system. When pre-training is not done, SI-SDRi shows a relative improvement of $34\%$ with the inclusion of real RIRs during training. When pre-training is done, a slightly lower relative improvement of $7\%$ of in SI-SDRi is observed. The relative improvements reported here are for training runs where RIRs from GAS are used.
    \item Curriculum learning improves the SI-SDRi by about $5\%$ relative to the non-curriculum baseline (averaged over all rooms). However, we see that curriculum learning does not improve performance when multi-stage training is applied (the last two rows of Table \ref{tab:ablation} show this). Applying curriculum learning in the later stages of multi-stage training scenario does not show the expected improvement as the model has already been exposed to RIRs for both high and low RT60 values in the first training stage.
    \item Pre-training the model on the non-reverberant separation task improves SI-SDRi by about $37\%$ relative to the non-pre-trained baseline (averaged over all rooms).
    \item Multi-stage training provides a relative improvement of $19.6\%$ over single stage training (averaged over all such runs in the above table over all room configurations).
    \item Fine-tuning on GAN RIRs gives a relative improvement of $16.7\%$ when compared to the GAS only training pipeline (averaged over all room configurations).
\end{itemize}

\section{Conclusion and Future Work}
\label{sec:conclusion}


We have shown that a combination of pre-training, multi-stage training and training on a combination of RIRs generated from accurate acoustic simulators and GAN based RIR generation methods can significantly improve the performance of reverberant speech separation compared to the ISM baseline. Our acoustic simulator generates RIRs based on specular and diffuse reflections, which considerably improve the accuracy. We hope to extend our approach to multi-channel and multi-speaker ($>2$ speakers) scenarios. These  techniques based on curriculum learning and accurate synthetic data generation could also be useful for audio-visual speech separation.




\bibliographystyle{IEEEbib}
\bibliography{strings,refs}

\begin{thebibliography}{10}

\bibitem{cooke2001auditory}
Martin Cooke and Daniel~PW Ellis,
\newblock ``The auditory organization of speech and other sources in listeners
  and computational models,''
\newblock {\em Speech communication}, vol. 35, no. 3-4, pp. 141--177, 2001.

\bibitem{schmidt2007speech}
Mikkel~N Schmidt,
\newblock ``Speech separation using non-negative features and sparse
  non-negative matrix factorization,''
\newblock in {\em Interspeech}. Citeseer, 2007, pp. 19--33.

\bibitem{hyvarinen2000independent}
Aapo Hyv{\"a}rinen and Erkki Oja,
\newblock ``Independent component analysis: algorithms and applications,''
\newblock {\em Neural networks}, vol. 13, no. 4-5, pp. 411--430, 2000.

\bibitem{hershey2016deep}
John~R Hershey, Zhuo Chen, Jonathan Le~Roux, and Shinji Watanabe,
\newblock ``Deep clustering: Discriminative embeddings for segmentation and
  separation,''
\newblock in {\em 2016 IEEE International Conference on Acoustics, Speech and
  Signal Processing (ICASSP)}. IEEE, 2016, pp. 31--35.

\bibitem{yu2017permutation}
Dong Yu, Morten Kolb{\ae}k, Zheng-Hua Tan, and Jesper Jensen,
\newblock ``Permutation invariant training of deep models for
  speaker-independent multi-talker speech separation,''
\newblock in {\em 2017 IEEE International Conference on Acoustics, Speech and
  Signal Processing (ICASSP)}. IEEE, 2017, pp. 241--245.

\bibitem{maciejewski2020whamr}
Matthew Maciejewski, Gordon Wichern, Emmett McQuinn, and Jonathan Le~Roux,
\newblock ``Whamr!: Noisy and reverberant single-channel speech separation,''
\newblock in {\em ICASSP 2020-2020 IEEE International Conference on Acoustics,
  Speech and Signal Processing (ICASSP)}. IEEE, 2020, pp. 696--700.

\bibitem{zeghidour2020wavesplit}
Neil Zeghidour and David Grangier,
\newblock ``Wavesplit: End-to-end speech separation by speaker clustering,''
\newblock {\em arXiv preprint arXiv:2002.08933}, 2020.

\bibitem{luo2021distortion}
Yi~Luo, Cong Han, and Nima Mesgarani,
\newblock ``Distortion-controlled training for end-to-end reverberant speech
  separation with auxiliary autoencoding loss,''
\newblock in {\em 2021 IEEE Spoken Language Technology Workshop (SLT)}. IEEE,
  2021, pp. 825--832.

\bibitem{aroudi2021dbnet}
Ali Aroudi and Sebastian Braun,
\newblock ``Dbnet: Doa-driven beamforming network for end-to-end reverberant
  sound source separation,''
\newblock in {\em ICASSP 2021-2021 IEEE International Conference on Acoustics,
  Speech and Signal Processing (ICASSP)}. IEEE, 2021, pp. 211--215.

\bibitem{zhang2021time}
Jisi Zhang, C{\u{a}}t{\u{a}}lin Zoril{\u{a}}, Rama Doddipatla, and Jon Barker,
\newblock ``Time-domain speech extraction with spatial information and multi
  speaker conditioning mechanism,''
\newblock in {\em ICASSP 2021-2021 IEEE International Conference on Acoustics,
  Speech and Signal Processing (ICASSP)}. IEEE, 2021, pp. 6084--6088.

\bibitem{bengio2009curriculum}
Yoshua Bengio, J{\'e}r{\^o}me Louradour, Ronan Collobert, and Jason Weston,
\newblock ``Curriculum learning,''
\newblock in {\em Proceedings of the 26th annual international conference on
  machine learning}, 2009, pp. 41--48.

\bibitem{braun2017curriculum}
Stefan Braun, Daniel Neil, and Shih-Chii Liu,
\newblock ``A curriculum learning method for improved noise robustness in
  automatic speech recognition,''
\newblock in {\em 2017 25th European Signal Processing Conference (EUSIPCO)}.
  IEEE, 2017, pp. 548--552.

\bibitem{ranjan2021curriculum}
Shivesh Ranjan and John~HL Hansen,
\newblock ``Curriculum learning based approaches for robust end-to-end
  far-field speech recognition,''
\newblock {\em Speech Communication}, 2021.

\bibitem{wang2020curriculum}
Chengyi Wang, Yu~Wu, Shujie Liu, Ming Zhou, and Zhenglu Yang,
\newblock ``Curriculum pre-training for end-to-end speech translation,''
\newblock {\em arXiv preprint arXiv:2004.10093}, 2020.

\bibitem{afouras2018deep}
Triantafyllos Afouras, Joon~Son Chung, Andrew Senior, Oriol Vinyals, and Andrew
  Zisserman,
\newblock ``Deep audio-visual speech recognition,''
\newblock {\em IEEE transactions on pattern analysis and machine intelligence},
  2018.

\bibitem{afouras2018conversation}
Triantafyllos Afouras, Joon~Son Chung, and Andrew Zisserman,
\newblock ``The conversation: Deep audio-visual speech enhancement,''
\newblock {\em arXiv preprint arXiv:1804.04121}, 2018.

\bibitem{hendrycks2019using}
Dan Hendrycks, Kimin Lee, and Mantas Mazeika,
\newblock ``Using pre-training can improve model robustness and uncertainty,''
\newblock in {\em International Conference on Machine Learning}. PMLR, 2019,
  pp. 2712--2721.

\bibitem{bansal2018pre}
Sameer Bansal, Herman Kamper, Karen Livescu, Adam Lopez, and Sharon Goldwater,
\newblock ``Pre-training on high-resource speech recognition improves
  low-resource speech-to-text translation,''
\newblock {\em arXiv preprint arXiv:1809.01431}, 2018.

\bibitem{zeyer2018improved}
Albert Zeyer, Kazuki Irie, Ralf Schl{\"u}ter, and Hermann Ney,
\newblock ``Improved training of end-to-end attention models for speech
  recognition,''
\newblock {\em arXiv preprint arXiv:1805.03294}, 2018.

\bibitem{schneider2019wav2vec}
Steffen Schneider, Alexei Baevski, Ronan Collobert, and Michael Auli,
\newblock ``wav2vec: Unsupervised pre-training for speech recognition,''
\newblock {\em arXiv preprint arXiv:1904.05862}, 2019.

\bibitem{huang2020self}
Sung-Feng Huang, Shun-Po Chuang, Da-Rong Liu, Yi-Chen Chen, Gene-Ping Yang, and
  Hung-yi Lee,
\newblock ``Self-supervised pre-training reduces label permutation instability
  of speech separation,''
\newblock {\em arXiv preprint arXiv:2010.15366}, 2020.

\bibitem{allen1979image}
Jont~B Allen and David~A Berkley,
\newblock ``Image method for efficiently simulating small-room acoustics,''
\newblock {\em The Journal of the Acoustical Society of America}, vol. 65, no.
  4, pp. 943--950, 1979.

\bibitem{diaz2021gpurir}
David Diaz-Guerra, Antonio Miguel, and Jose~R Beltran,
\newblock ``gpurir: A python library for room impulse response simulation with
  gpu acceleration,''
\newblock {\em Multimedia Tools and Applications}, vol. 80, no. 4, pp.
  5653--5671, 2021.

\bibitem{cosentino2020librimix}
Joris Cosentino, Manuel Pariente, Samuele Cornell, Antoine Deleforge, and
  Emmanuel Vincent,
\newblock ``Librimix: An open-source dataset for generalizable speech
  separation,''
\newblock {\em arXiv preprint arXiv:2005.11262}, 2020.

\bibitem{panayotov2015librispeech}
Vassil Panayotov, Guoguo Chen, Daniel Povey, and Sanjeev Khudanpur,
\newblock ``Librispeech: an asr corpus based on public domain audio books,''
\newblock in {\em 2015 IEEE international conference on acoustics, speech and
  signal processing (ICASSP)}. IEEE, 2015, pp. 5206--5210.

\bibitem{szoke2019building}
Igor Sz{\"o}ke, Miroslav Sk{\'a}cel, Ladislav Mo{\v{s}}ner, Jakub Paliesek, and
  Jan {\v{C}}ernock{\`y},
\newblock ``Building and evaluation of a real room impulse response dataset,''
\newblock {\em IEEE Journal of Selected Topics in Signal Processing}, vol. 13,
  no. 4, pp. 863--876, 2019.

\bibitem{richey2018voices}
Colleen Richey, Maria~A. Barrios, Zeb Armstrong, Chris Bartels, Horacio Franco,
  Martin Graciarena, Aaron Lawson, Mahesh~Kumar Nandwana, Allen Stauffer,
  Julien van Hout, Paul Gamble, Jeff Hetherly, Cory Stephenson, and Karl Ni,
\newblock ``Voices obscured in complex environmental settings (voices)
  corpus,'' 2018.

\bibitem{ko2017study}
Tom Ko, Vijayaditya Peddinti, Daniel Povey, Michael~L Seltzer, and Sanjeev
  Khudanpur,
\newblock ``A study on data augmentation of reverberant speech for robust
  speech recognition,''
\newblock in {\em 2017 IEEE International Conference on Acoustics, Speech and
  Signal Processing (ICASSP)}. IEEE, 2017, pp. 5220--5224.

\bibitem{hodgson1991evidence}
Murray Hodgson,
\newblock ``Evidence of diffuse surface reflections in rooms,''
\newblock {\em The Journal of the Acoustical Society of America}, vol. 89, no.
  2, pp. 765--771, 1991.

\bibitem{dalenback1994macroscopic}
Bengt-Inge Dalenb{\"a}ck, Mendel Kleiner, and Peter Svensson,
\newblock ``A macroscopic view of diffuse reflection,''
\newblock {\em Journal of the Audio Engineering Society}, vol. 42, no. 10, pp.
  793--807, 1994.

\bibitem{krokstad1968}
Asbj{\o}rn Krokstad, S~Strom, and Svein S{\o}rsdal,
\newblock ``Calculating the acoustical room response by the use of a ray
  tracing technique,''
\newblock {\em Journal of Sound and Vibration}, vol. 8, no. 1, pp. 118--125,
  1968.

\bibitem{lauterbach2007}
Christian Lauterbach, Anish Chandak, and Dinesh Manocha,
\newblock ``Interactive sound rendering in complex and dynamic scenes using
  frustum tracing,''
\newblock {\em IEEE Transactions on Visualization and Computer Graphics}, vol.
  13, no. 6, pp. 1672--1679, 2007.

\bibitem{savioja2015overview}
Lauri Savioja and U~Peter Svensson,
\newblock ``Overview of geometrical room acoustic modeling techniques,''
\newblock {\em The Journal of the Acoustical Society of America}, vol. 138, no.
  2, pp. 708--730, 2015.

\bibitem{schissler2014high}
Carl Schissler, Ravish Mehra, and Dinesh Manocha,
\newblock ``High-order diffraction and diffuse reflections for interactive
  sound propagation in large environments,''
\newblock {\em ACM Transactions on Graphics (TOG)}, vol. 33, no. 4, pp. 39,
  2014.

\bibitem{schissler2018interactive}
Carl Schissler and Dinesh Manocha,
\newblock ``Interactive sound rendering on mobile devices using
  ray-parameterized reverberation filters,''
\newblock {\em arXiv preprint arXiv:1803.00430}, 2018.

\bibitem{tang2020improving}
Zhenyu Tang, Lianwu Chen, Bo~Wu, Dong Yu, and Dinesh Manocha,
\newblock ``Improving reverberant speech training using diffuse acoustic
  simulation,''
\newblock in {\em ICASSP 2020-2020 IEEE International Conference on Acoustics,
  Speech and Signal Processing (ICASSP)}. IEEE, 2020, pp. 6969--6973.

\bibitem{ratnarajah2020ir}
Anton Ratnarajah, Zhenyu Tang, and Dinesh Manocha,
\newblock ``Ir-gan: Room impulse response generator for speech augmentation,''
\newblock {\em arXiv preprint arXiv:2010.13219}, 2020.

\bibitem{ratnarajah2021ts}
Anton Ratnarajah, Zhenyu Tang, and Dinesh Manocha,
\newblock ``Ts-rir: Translated synthetic room impulse responses for speech
  augmentation,''
\newblock {\em arXiv preprint arXiv:2103.16804}, 2021.

\bibitem{luo2020dual}
Yi~Luo, Zhuo Chen, and Takuya Yoshioka,
\newblock ``Dual-path rnn: efficient long sequence modeling for time-domain
  single-channel speech separation,''
\newblock in {\em ICASSP 2020-2020 IEEE International Conference on Acoustics,
  Speech and Signal Processing (ICASSP)}. IEEE, 2020, pp. 46--50.

\bibitem{kingma2014adam}
Diederik~P Kingma and Jimmy Ba,
\newblock ``Adam: A method for stochastic optimization,''
\newblock {\em arXiv preprint arXiv:1412.6980}, 2014.

\bibitem{pariente2020asteroid}
Manuel Pariente, Samuele Cornell, Joris Cosentino, Sunit Sivasankaran,
  Efthymios Tzinis, Jens Heitkaemper, Michel Olvera, Fabian-Robert St{\"o}ter,
  Mathieu Hu, Juan~M Mart{\'\i}n-Do{\~n}as, et~al.,
\newblock ``Asteroid: the pytorch-based audio source separation toolkit for
  researchers,''
\newblock {\em arXiv preprint arXiv:2005.04132}, 2020.

\bibitem{raffel2014mir_eval}
Colin Raffel, Brian McFee, Eric~J Humphrey, Justin Salamon, Oriol Nieto, Dawen
  Liang, Daniel~PW Ellis, and C~Colin Raffel,
\newblock ``mir\_eval: A transparent implementation of common mir metrics,''
\newblock in {\em ISMIR}. Citeseer, 2014.

\end{thebibliography}

\end{document}